\documentclass{aa}
\usepackage{graphicx}

\begin{document}
\newcommand{\MIC}{\mbox{$\mu$m}}
\newcommand{\NiII}{\mbox{[Ni$\,$II]}}
\newcommand{\PII}{\mbox{[P$\,$II]}}
\newcommand{\FeII}{\mbox{[Fe$\,$II]}}
\newcommand{\FeXIII}{\mbox{[Fe$\,$XIII]}}
\newcommand{\FeXIV}{\mbox{[Fe$\,$XIV]}}
\newcommand{\SII}{\mbox{[S$\,$II]}}
\newcommand{\OI}{\mbox{[O$\,$I]}}
\newcommand{\Feplus}{\mbox{Fe$^+$}}
\newcommand{\Pplus}{\mbox{P$^+$}}
\newcommand{\EW}{\mbox{$W_\lambda$}}
\newcommand{\LSUN}{\mbox{L$_\odot$}}

\title{
NICS--TNG infrared spectroscopy of NGC1068:
the first extragalactic measurement of \PII\  and a new tool to constrain
the origin of \FeII\  line emission in galaxies.
}

\titlerunning{
%NICS--TNG discovery of \PII\  in NGC1068
%The first extragalactic detection of \PII\  by NICS--TNG
%Infrared \PII\  lines and the origin of \FeII\  in AGN's
[PII] in NGC1068 and the role of shocks in AGNs
}
\authorrunning{E. Oliva, A. Marconi, R. Maiolino et al.}

\author
{
E. Oliva\inst{1,2},
A. Marconi\inst{1},
R. Maiolino\inst{1},
L. Testi\inst{1},
F. Mannucci\inst{3},
F. Ghinassi\inst{2},
J. Licandro\inst{2},
L. Origlia\inst{4},  % Per il lavoro che ha fatto con IRSPEC sui SNRs
C. Baffa\inst{1},
A. Checcucci\inst{1},
G. Comoretto\inst{1},
V. Gavryussev\inst{3},
S. Gennari\inst{1},
E. Giani\inst{1},
L.K. Hunt\inst{3},
F. Lisi\inst{1},
D. Lorenzetti\inst{5},
G. Marcucci\inst{6},
L. Miglietta\inst{1},
M. Sozzi\inst{3},
P. Stefanini\inst{1},
F. Vitali\inst{5}
}
          
\institute{ 
Osservatorio di Arcetri, Largo E. Fermi 5, I-50125 Firenze, Italy
\and
Centro Galileo Galilei \& Telescopio Nazionale Galileo, P.O. Box 565
E-38700 S. Cruz de La Palma, Spain
\and
CAISMI-CNR,  Largo E. Fermi 5, I-50125 Firenze, Italy
\and
Osservatorio Astronomico di Bologna, Via Ranzani 1, I-40127 Bologna, Italy
\and
Osservatorio Astronomico di Roma, Via Frascati 33, I-00044 Rome Italy
\and
Universit\`a degli studi di Firenze, dipartimento di Astronomia,
 Largo E. Fermi 5, I-50125 Firenze, Italy
}

\offprints{E. Oliva, e-mail oliva@tng.iac.es }

%\thesaurus{ 03 ( 
%               02.12.1;          % Line: formation
%               02.12.2;          % Line: identification
%               11.01.2;          % Galaxies: active
%               11.09.1;          % Galaxies: individual: NGC1068
%               11.19.1;          % Galaxies: Seyfert
%               13.09.1          % Infrared: galaxies
%               )
%          }

\date{ Received  26 January 2001 / Accepted 7 February 2001 }

\abstract{ 
We report 0.9-1.4 \MIC\  spectroscopic observations of 
NGC1068 collected during the commissioning phase of the
near infrared camera spectrometer (NICS) of the Telescopio
Nazionale Galileo (TNG).
These yielded the first extragalactic measurement of \PII\  (1.188 \MIC)
line emission. 
In the central 0.75"x2" the \FeII/\PII\    
line-intensity ratio is close to unity, 
similar to that measured in the Orion bar and
a factor of $\la$20 smaller than in supernova remnants.
This indicates that most of iron is locked
into grains and, therefore, argues against shock excitation being the
primary origin of \FeII\  line emission in the central regions of NGC1068.\\
We propose the \FeII/\PII\  ratio as a simple and effective tool to
study and perhaps resolve the long debated questions related to the
origin of \FeII\  line emission and, more generally, 
to constrain the role of shock excitation in active galaxies. 
}
\maketitle

\begin{figure*}
\centering
\resizebox{\hsize}{!} {\rotatebox{-90}{\includegraphics{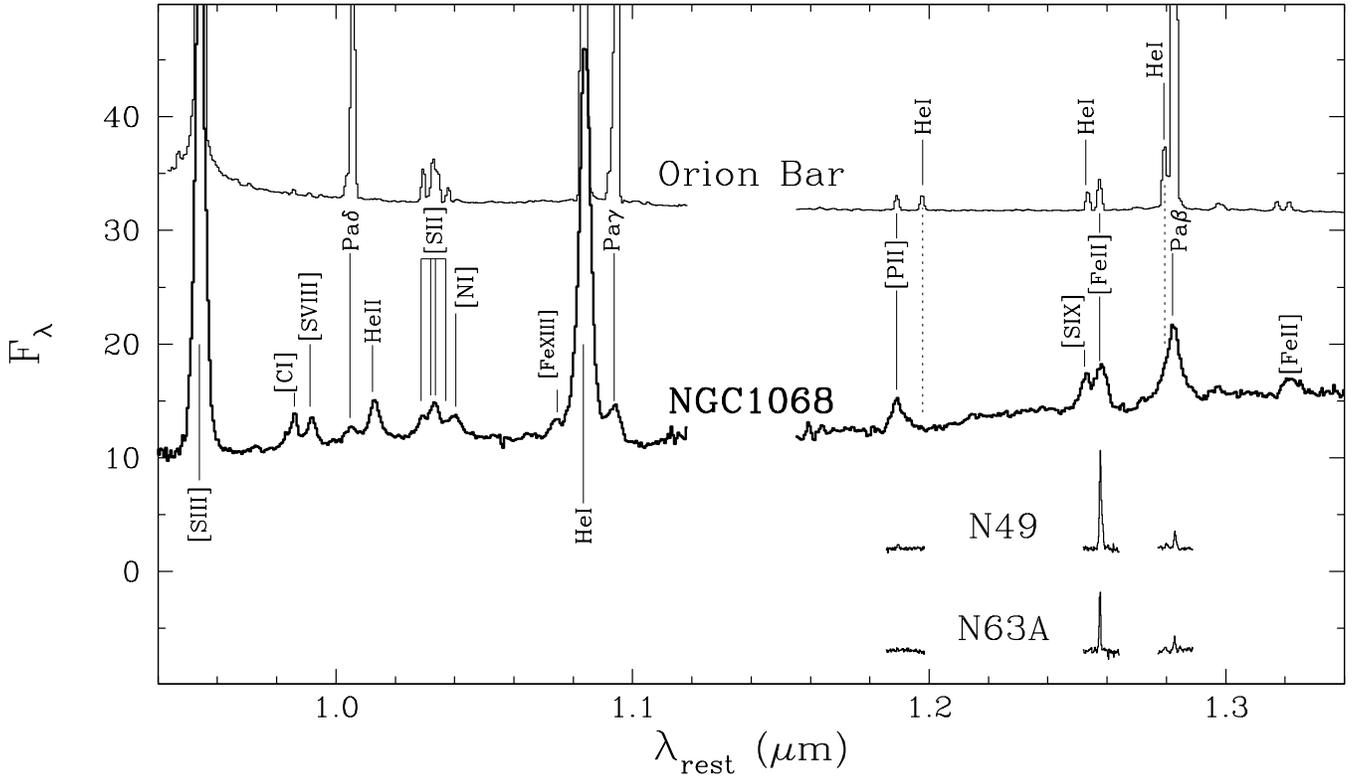}}}
%\sidecaption
%\resizebox{12cm}{!} {\rotatebox{-90}{\includegraphics{Da262_F1.ps}}}
%\includegraphics[width=10cm,angle=-90]{Da262_f1.ps}
\caption{
NICS--TNG spectrum of the Seyfert galaxy
NGC1068 compared with spectra of template 
objects from the literature (see text Sect.~\ref{results}). 
The break at $\simeq$1.1 \MIC\  corresponds to the region of bad
atmospheric transmission.
The flux scales
and zero levels have been adjusted to facilitate the direct comparison
between the spectra.
}
\label{fig_spec}
\end{figure*}

\section{Introduction} 
\label{introduction}
Since the first infrared spectroscopic observations of galaxies and
supernova remnants
in the 80's, the emission lines of \FeII\  have become a popular and
debated issue (Moorwood \& Oliva \cite{moorwood}, 
Forbes \& Ward \cite{forbes93},
Simpson et al. \cite{simpson96},
Veilleux et al. \cite{veilleux97},
Alonso--Herrero et al. \cite{alonso97},
Mouri et al. \cite{mouri00}).
  
From the observational point of view, \FeII\  is weak
in HII regions and planetary nebulae while extremely strong in
shock--excited filaments of supernova remnants.
Since relatively bright \FeII\  emission is commonly found 
in the IR spectra of normal and active 
galaxies, many
authors have considered the possibility
of using this line as shock tracer and, even, to count the number 
of supernova remnants (e.g. Colina \cite{colina93},
Vanzi \& Rieke \cite{vanzi97},
Engelbracht et al. \cite{engelbracht98}).

From the theoretical point of view, a low density region with normal abundances
can become a strong source of \FeII\  only if the following conditions
are satisfied.
\begin{itemize}
\item[$i)$] Most of iron must be in the gas phase, i.e. 
dust grains must have been destroyed.
\item[$ii)$] The gas electron temperature must be large enough to 
collisionally excite the upper levels of the lines, in practice 
$T_e\!\ga\!5000$ K. 
\item[$iii)$] Most of iron must be singly ionized, i.e. \Feplus/Fe must
be close to unity.
\end{itemize}
Given the low ionization potential of \Feplus\  and the high efficiency of
the Fe$^{++}$ + H$^o$ charge--exchange recombination reactions, 
the latter two conditions are equivalent to saying that bright \FeII\  
lines can only be formed in regions where hydrogen is partly
ionized (e.g. Oliva et al. \cite{oliva_snr1}). 

The most efficient 
mechanisms for creating extended regions of hot, partially 
ionized gas are shocks and photoionization by soft X--rays. In both
cases the volume emission measure of the partially ionized region
could easily exceed that of the fully ionized gas. The only
important difference between the two mechanisms is that photoionization
is unable to destroy the toughest iron--based grains which are otherwise
easily sputtered by shock fronts.

In practice, therefore, a purely photoionized region emitting \FeII\ 
can be easily recognized by measuring the abundance of \Feplus\  relative 
to any non-refractory species which forms in the same partially ionized
region. Finding a very low iron abundance would unequivocally imply that the
gas has not been significantly processed by shocks.

This could be in principle obtained by comparing the infrared \FeII\
and optical \OI\  lines but, in practice, the large difference in 
critical densities, reddening
and the problems of comparing data taken with different instruments 
makes it impossible to obtain a clear--cut conclusion (see e.g. 
Mouri et al. \cite{mouri93},
Alonso--Herrero et al. \cite{alonso97},
Larkin et al. \cite{larkin98}).
A much more reliable determination of the iron relative abundance should
be derived from lines close in wavelengths and with
similar critical densities such as the
\PII\  and \FeII\  lines which are discussed in this
Letter.

\section{Observations and results}
\label{results}
\begin{table*}
\caption{Line fluxes in NGC1068 and template objects}

\newcommand{\SKIP}{\noalign{\vskip2pt}}
\begin{flushleft}
%\begin{tabular}{lrrrrr}
\begin{tabular}{lccccc}
\hline\hline
\SKIP
Line ($\mu$m)   &   \multicolumn{5}{c}{Relative fluxes$^{(1)}$} \\
%  Orion: position A of Walmsley et al.
%  RCW103: Oliva et al. 1990
%
                & NGC1068 & Orion & RCW103 & LMC-N63A & LMC-N49 \\
%                &  &  & \multicolumn{3}{c}{(supernova remnants)} \\
\SKIP
\hline
\SKIP
\ [SIII] 0.9529   &  1200   & 29000 &   --   &    --   &  --  \\
\ [CI] 0.985      &    72   &   20  &   --   &    --   &  --  \\
\ [SVIII] 0.9913  &    55   & $<$10 &   --   &    --   &  --  \\
\ Pa$\delta$ 1.005 &   61:  & 1400  &   --   &    --   &  --  \\
\ HeII 1.012      &   120   & $<$10 &   --   &    --   &  --  \\
\ [SII] 1.033     &   170:  &  400  &   --   &    --   &  --  \\ 
\ [NI] 1.040      &    60:  &   14  &   --   &    --   &  --  \\ 
\ [FeXIII] 1.075  &   50:  & $<$10 &   --   &    --   &  --  \\
\ HeI 1.083       &  1000   & 6600  &   31   &    --   &  --  \\
\ Pa$\gamma$ 1.094 &  110:  & 2300  & $<$15  &    --   &  --  \\
\ [PII] 1.188     &    67   &   50  & $<$8   &  $<$6   &   3  \\
\ HeI 1.197       &  $<$20 &   48  & $<$8   &  $<$6   &  $<$6 \\
\ [SIX] 1.252 + HeI 1.253 
                 & 72$^a$ & 66$^b$ & $<$8 &  $<$6 & $<$6  \\
\ [FeII] 1.257    &   100   &  100  &  100   &   100   & 100  \\
\ HeI 1.279       &    50:  &  430  &   --   &  $<$6   &  $<$6   \\
\ Pa$\beta$ 1.282 &   230   & 4300  &   11   &    20   &  15  \\
\ [FeII] 1.321    &    33   &   34  &   30   &    --   &  --  \\
\hline
\SKIP
\ [FeII] intensity$^{(2)}$ & $\approx$20$^c$ & 4.4  &   12  &   3.0 &   4.4 \\
\SKIP
\hline\hline
\end{tabular}
\begin{enumerate}
\item[$^{(1)}$] Normalized to I([FeII] 1.257)=100.
Fluxes for Orion refer to position A of Walmsley et al. 
(\cite{orion2000}),
values for the supernova remnants are from Oliva et al. 
(\cite{rcw103}) and from
the IRSPEC spectra displayed in Fig.~\ref{fig_spec}.
The error on line fluxes in NGC1068 are tyipically
$\pm$10\% except for the entries
marked with a ``:'' which are uncertain due to blending.
\item[$^{(2)}$] Intensity of [FeII] 1.257 in units of $10^{-4}$ erg cm$^{-2}$
s$^{-1}$ sr$^{-1}$
\item[$^a$] Contribution from HeI 1.253 should be $\simeq$10, 
based on the observed intensities of the other HeI lines.
\item[$^b$] Contribution from [SIX] is negligible.
\item[$^c$] The absolute flux calibration is uncertain.
%\end{list} 
\end{enumerate} 
\end{flushleft}
\label{tab_fluxes}
\end{table*}

The data were collected at the Telescopio Nazionale Galileo (TNG)
in November 2000 during the commissioning phase of NICS, the near infrared 
camera and spectrometer expressly designed and built for this telescope.
This instrument is a FOSC--type cryogenic focal reducer equipped 
with two interchangeable cameras feeding a Rockwell Hawaii 1024$^2$ array. 
The camera used for the spectroscopic observations has 
a projected scale of 0.25"/pixel
(Oliva \& Gennari \cite{gennari1},  
Baffa et al. \cite{baffa}).
The spectroscopic modes are achieved by means of a series of glass--resin
grisms which can be inserted in the 22 mm collimated beam
(Vitali et al. \cite{vitali}).
The spectrum of NGC1068 was collected through a slit of 0.75" (=3 pixels)
width and using the IJ grism which 
%consists of a 36.8$^o$ prism of irgn6 glass with a
%resin replica of a 300 lines/mm, 36.8$^o$ blaze grating and a
%RG850 coloured glass as order sorter. This mode
yields a 0.89-1.46 \MIC\
spectrum with a dispersion of 5.7 \AA/pix.
% equivalent, in our case, to a resolving power $R$=700 at 1.2 \MIC.
%
The slit was oriented N--S (i.e. at PA=0$^\circ$) and centered
on the 1 $\mu$m continuum peak. The acquisition consisted
of a series of four 5--minute exposures 
with the object set at different positions along the slit followed
by halogen flats.
The atmospheric spectral response and the instrumental 
efficiency were determined
using spectra of the 
O6.5V star HD42088 whose intrinsic spectrum
was approximated by $F_\lambda$=1.7$\cdot 10^{-9}$ 
($\lambda$/1.25)$^{-3.7}$ erg cm$^{-2}$ s$^{-1}$ 
$\mu$m$^{-1}$.

The spectrum of the central 0.75"x2" region is displayed in Fig.~\ref{fig_spec}
where we also show, for comparison, spectra of the Orion Bar (Walmsley
et al. 2000) and unpublished spectra of supernova remnants
collected in 1992 using IRSPEC at the ESO-NTT telescope.
The relative line fluxes are summarized in Table~\ref{tab_fluxes}.
Evident is the difference between the very large \FeII/\PII\  ratio measured
in SNR's and the much smaller values found in the Orion Bar and in NGC1068.

In principle, the emission feature peaking at 1.188 $\mu$m could
be contaminated by \NiII\ 1.191 $\mu$m. %  which lies very nearby to \PII.
However, this  line was measured in the Crab nebula at a level of
only $\la$15\% of \FeII\  (Rudy et al.  \cite{rudy}) and was not
detected in the supernova remnants listed in Table~\ref{tab_fluxes}.
Moreover,  \NiII\  has a critical density of only
$\simeq$500 cm$^{-3}$ 
(Nusbbaumer \& Storey \cite{nussbaumer2}),
two orders of magnitude lower than the critical density of \FeII\  and 
much lower than the electron density pertinent to the region of NGC1068
under consideration here. For these reasons we believe that \NiII\
does not affect significantly the measurement of \PII. 
%A more direct and definite answer could be obtained by measuring
%the other \PII\  line at 1.147 $\mu$m line which arises from the same upper
%level as \PII\  1.188 but lies, unfortunately, in a region of poor
%atmospheric transmission.

It is also interesting to note that our data 
confirm the identification of [SIX] by Marconi et al.
(\cite{marconi96}) and include the first detection of \FeXIII\   in
an extragalactic object. 
The latter identification is also supported by 
measurements of the higher ionization green line 
of \FeXIV\  (Kraemer \& Crenshaw \cite{kraemer}).

\section{Discussion}

\subsection{ \FeII, \PII\  and the Fe/P abundance ratio }
\label{discussion1}

The near--IR lines of \PII\  and \FeII\  have several interesting
similarities. They lie nearby in wavelength, 
have similar excitation temperatures and critical densities 
and their parent ions have similar ionization potentials and
radiative recombination coefficients.
Using the collision strengths of
Krueger \& Czyzak (\cite{krueger}), 
Zhang \& Pradhan (\cite{zhang})
 and the transition probabilities of
Mendoza \& Zeippen (\cite{mendoza}),
Nussbaumer \& Storey (\cite{nussbaumer})
yields
$$ {n(\Feplus)\over n(\Pplus)} \simeq 
         2 \cdot {I(\FeII\ 1.257 \MIC) \over I(\PII\ 1.188 \MIC)} \eqno(1)$$
which is accurate within a factor of 2 for all the temperatures and 
densities of interest.
The only important difference between the two species
is that charge exchange recombinations between P$^{++}$
and neutral hydrogen are $\simeq$2 orders of magnitude
less efficient than the Fe$^{++}$ + H$^o$
reactions (Kingdon \& Ferland \cite{kingdon96}). This implies 
$$ {n(\Feplus)\over n(\Pplus)} \ga 
           {n({\rm Fe}) \over n({\rm P}) } \eqno(2)$$
which combined with Eq. (1) yields
$$ {n({\rm Fe}) \over n({\rm P}) } \la 
      2 \cdot {I(\FeII\ 1.257 \MIC) \over I(\PII\ 1.188 \MIC)} \eqno(3)$$
i.e. the Fe/P abundance ratio is quite well constrained by the \FeII/\PII\ 
ratio and, if anything, it is overestimated.
Finally, it is interesting to note that, for a solar Fe/P$\simeq$100
abundance ratio, one expects \FeII/\PII$\simeq$50
i.e. a ratio similar
to that measured in supernova remnants.

\subsection{ Fe/P abundance ratio and dust destruction}
\label{discussion2}

Iron is a well known refractory species whose gas phase abundance 
in the ISM is
often found to be down by many orders of magnitudes relative
its cosmic value. % (e.g. Walmsley \cite{malcolm}). 
The only regions where
iron is not found to be significantly depleted are those associated 
with fast ($\ga$100 km/s) shocks which can effectively
destroy the grains by sputtering.
In normal photoionized regions the depletion of iron ranges between
the factor of $\simeq$0.1 measured in the Orion Bar 
(Baldwin et al. \cite{baldwin}) to significantly lower values
found in planetary nebulae (e.g. Oliva et al. \cite{oliva6302},
Perinotto et al. \cite{perinotto}). 
These differences probably reflect the fact that a variable
(though small) fraction of iron is locked into
relatively soft grains which can be easily destroyed without the need
of fast shocks.

Phosphorus is a non-refractory species whose measured depletion 
in ionized gas is close to unity. 
Therefore, the Fe/P relative abundance should give a direct estimate
of the iron depletion or, equivalently, of the presence of robust dust in
a given region.
This is indeed confirmed by the data presented here (Fig.~\ref{fig_spec} and
Table~\ref{tab_fluxes}) 
which show variations by more than one order of magnitude
between the large, quasi--solar Fe/P ratio found in SNR's and the
much smaller values found in the other objects.

\subsection{ \FeII/\PII\  and the origin of \FeII\  in galaxies}
\label{discussion3}

As already discussed in the introduction, determining the origin of \FeII\
line emission is of crucial importance for any 
program aiming at using \FeII\  for tracing shock fronts and/or
constraining the supernovae rate in galaxies.
The results obtained here indicate that the \FeII/\PII\  ratio could
provide a clear--cut answer to this problem.
 The line ratio is large ($\ga$20) in regions excited by
fast shocks while low ($\la$2) in normal photoionized region and in NGC1068.
This indicate that shocks are {\it not}
the dominant source of \FeII\  in the central $\simeq$200 pc 
of NGC1068 where,
most likely, the lines are produced by photoionization from the active
nucleus, as already indicated by detailed photoionization modeling
(e.g. Kraemer \& Crenshaw \cite{kraemer}).
The copious flux of soft X--rays from the AGN creates a very extended
partially ionized region which is responsible for the strong emission
of low ionization species such as \SII, \OI, \PII\ and \FeII.
However, the latter is relatively weak because most of the iron 
is locked into dust grains.
The relative intensities of the low and higher ionization lines depend
on a complex combination of the ionization parameter, density and of
the spectral shape of the ionizing radiation. Nevertheless, the ratio
between ``close relatives'' such as \FeII/\PII\  are little
influenced by this and, as discussed in Sect.~\ref{discussion1},
almost solely depend on the Fe/P relative abundance which, in turn,
is a direct measurement of iron depletion (Sect.~\ref{discussion2}).

\section{Conclusions}

Given the above results and considerations, we propose using the
ratio between \FeII\  (1.257 \MIC) and \PII\  (1.188 \MIC)
as a general tool for constraining the origin of
\FeII\  line emission in galaxies. 
%These lines lie close in
%wavelength and remain in a region of
%good atmospheric transmission for objects up to a redshift of 
%$z\!\simeq\!0.06$.\\
The diagnostic works as follows
\begin{itemize}

\item[--] In objects with low \FeII/\PII\  ratios  shocks do {\it not} 
play an important role in the lines excitation.

\item[--] Large values of \FeII/\PII\  ($\ga$20) indicate that the emitting gas
has recently passed through a fast shock which sputtered and
destroyed most of the dust grains. It is therefore likely that the
lines are also produced by shock excitation
%, though some contribution from photoionization cannot be excluded.
\end{itemize}

\begin{acknowledgements}
This paper is
based on observations made
  with the Italian Telescopio Nazionale Galileo (TNG) 
operated on the island of La Palma by the Centro Galileo
 Galilei of the CNAA (Consorzio Nazionale per l'Astronomia e l'Astrofisica) 
at the Spanish Observatorio del
Roque de los Muchachos of the Instituto de Astrofisica de Canarias.
We are grateful to all the technical staff and telescope operators for
their assistance during the commissioning phase of NICS.
\end{acknowledgements}

\end{document}